\pdfoutput=1
\RequirePackage{ifpdf}
\documentclass{JINST}
\usepackage{amsmath}

\author{Z. Moss, M. Toups, L. Bugel, G.H. Collin, J.M. Conrad\\
Department of Physics, Massachusetts Institute of Technology\\
E-mail: \email{mtoups@mit.edu}
}
\title{Anode-Coupled Readout for Light Collection in Liquid Argon TPCs}

\abstract{
This paper will discuss a new method of signal read-out from photon detectors in ultra-large, underground liquid argon time projection chambers.     
In this design, the signal from the light collection system is coupled via capacitive plates to the TPC wire-planes. This signal is then read out using the same cabling and electronics as the charge information.    %
This greatly benefits light collection: it eliminates the need for an independent readout, substantially reducing cost; It reduces the number of cables in the vapor region of the TPC that can produce impurities; And it cuts down on the number of feed-throughs in the cryostat wall that can cause heat-leaks and potential points of failure. We present experimental results that demonstrate the sensitivity of a LArTPC wire plane to photon detector signals. We also simulate the effect of a 1 $\mu$s shaping time and a 2 MHz sampling rate on these signals in the presence of noise, and find that a single photoelectron timing resolution of $\sim$30 ns can be achieved.
}

\begin{document}
\maketitle

\section{Introduction}

Efficient light collection systems are required for future ultra-large, underground LArTPCs in order to trigger and to establish the interaction-time of each event ($t_0$). This is especially important for the reconstruction of non-beam events where the $t_0$ is not dictated by the spill-time of an accelerator beam.
The light collection scheme for the Deep Underground Neutrino Experiment (DUNE) \cite{DUNE} provides a typical example of such a system. This 40 kT LArTPC is expected to deploy tens of thousands of silicon photo-multipliers (SiPMs) that will register and readout the light signal collected from flat-panel light-guides situated in the detector volume \cite{DUNEdesign}. The signals from the SiPMs will be read out through cables that extend from below the liquid argon level, through the gas-filled ullage above the liquid, and to the feed-throughs. On the warm side of the cryostat, the signals are brought from the feed-throughs to a dedicated readout for processing.

\par
The cables, feed-throughs, and readout electronics associated with light collection introduce logistical problems, potential points of failure, and considerable cost into a design like DUNE's.  Not only will the cables require routing that is especially complicated in the installation of the DUNE light collection system (where the photo-detectors are inserted between wrapped wire planes), but those cables will also extend through the gas-filled ullage. Cables can introduce impurities into the ullage \cite{Pordes}, thereby placing a burden on the purification system.  The signal cable feed-throughs themselves are also complex and constitute potential points of failure. Large feed-throughs additionally allow heat to leak into the cryostat.  Finally, the cables, connectors, high density signal feed-throughs, and readout channels all contribute to the overall cost of the light collection system. Costs for ADC readout systems for photomultiplier tubes and SiPMs are typically several hundred dollars per channel, and represent a significant fraction of the total cost of the light collection system in an ultra-large detector.  Overall, the problems associated with a dedicated optical readout system may limit the total light collection coverage that can be practically installed in an ultra-large detector.   
\par

For these reasons, we sought a method to replace the dedicated optical readout for ultra-large LArTPCs altogether.  We propose to couple the photodetector signals capacitively to the wires of the TPC anode plane via conductive plates. This system is free of physical connection to the wire plane and associated electronics, simplifying logistics during installation. In contrast to a single-capacitor coupling to the wire-plane, this system has the advantage of alleviating issues related to noise by spreading the light collection signal across adjacent wires in a characteristic manner.

\par Use of the wire plane for both light and charge signals naturally raises concerns over signal overlap, that can complicate event reconstruction.  However, for a given energy deposition in the fiducial volume of a large LArTPC detector, 
the induced light signals appear in the TPC readout well before the charge signals due to the relatively slow electron drift time of $\sim$1mm/$\mu$s \cite{uBTDR}, \cite{sbn}.  In deep underground neutrino LArTPC experiments, such as DUNE, it is also unlikely that those charge signals will be contaminated by other induced light signals from cosmic muons arriving during the TPC readout window (the estimated cosmic ray rate throughout the DUNE detector is only 0.26 Hz \cite{DUNEdesign}). Because the induced signals produce a distinctive pattern spread over several TPC wires, it should also be possible to disentangle and subtract them from the charge signals.  We will show that the plate system works best when the plate is in close proximity to the wires.  This coordinates well with designs of detectors such as DUNE, where the light collection system is to be installed in the centimeter scale gaps between adjacent collection wire planes. If 6 mm $\times$ 6 mm SensL SiPMs (e.g. MicroFC-60035-SMT) are used for photo-detection, then cables would only be required for the $\sim$25 V bias voltage, which can be delivered on a bus that powers many devices at once.

\begin{figure}[t]
	\centering
    \begin{minipage}[b]{1.0\textwidth}
    \centering
      \begin{tabular}{c}
\includegraphics[width=0.5\columnwidth]{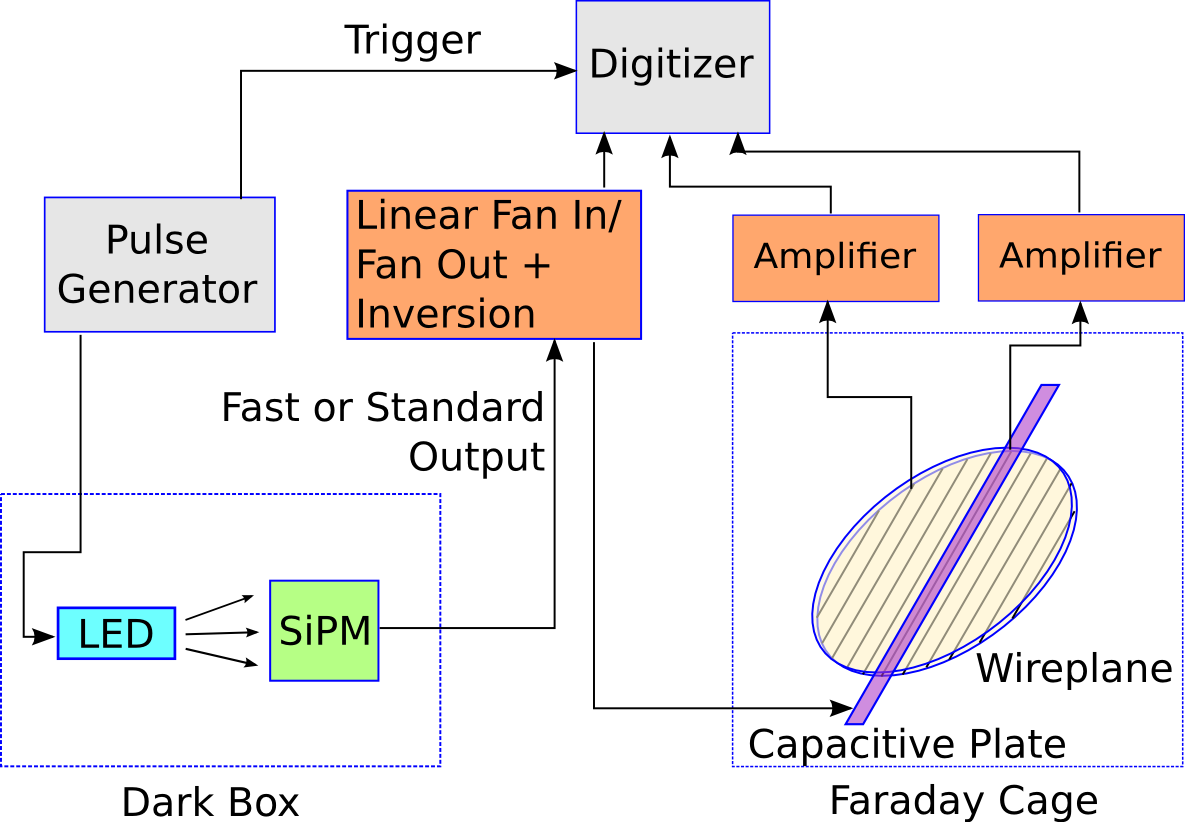}\\
\vspace{0.25in}\\
\includegraphics[width=0.5\columnwidth]{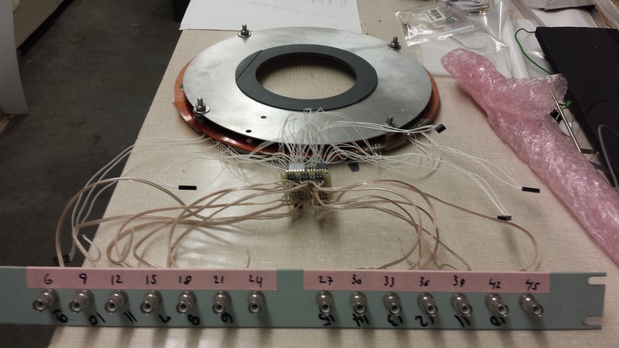}\\ 
				\end{tabular}
      \caption{The experimental apparatus;  Top: A schematic diagram; Bottom: A photograph of the wire plane and patch panel for readout.}\label{fig:wireplane}
    \end{minipage}%
 
\end{figure}

\par	
	In what follows, we demonstrate the viability of our capacitively-coupled readout design.  As a proof of principle, we use signals from the SiPMs proposed for DUNE to induce signals on a small LArTPC wire-plane via a copper strip suspended near the wires. We proceed to characterize the spatial distribution of induced signal amplitudes over the plane, as well as the linearity of the wire plane response. Measurements of a linear response can be used to estimate the sensitivity of such a system to single photoelectrons (SPEs).  We then simulate the effect of shaping and digitization of these wire-plane signals using realistic assumptions based on the wire-plane readout of the MicroBooNE LArTPC at Fermilab \cite{uB, uBTDR}.  We show that despite the relatively slow sampling speed of the TPC electronics (by comparison to dedicated optical readout), single photoelectron timing resolutions of $\sim$30 ns can be achieved for a single 6 mm $\times$ 6 mm SensL SiPM signal coupled via a strip of copper to the wires.

\section{Wire-plane Experiments \label{Experiments}}

Before studying the capabilities of such a system in detail, we first demonstrated, in hardware, the sensitivity of a LArTPC wire-plane to unamplified SiPM signals. The experimental apparatus, diagrammed in Figure~\ref{fig:wireplane} (top) consisted of a plane of 50 125 $\mu$m diameter CuBe wires, spaced at 4.7 mm intervals, with 14 wires instrumented (shown in Figure~\ref{fig:wireplane}, bottom).   The wire plane used in this study was originally part of a small TPC already used in liquid argon R\&D~\cite{longbo}.  Every third wire was instrumented in order to provide broad spatial coverage of the wire-plane, given only 14 channels of readout. We determined that the SiPM itself does not induce significant signal on the wire plane, and so a capacitive plate is required to increase the coupling between the SiPM signal and the wire-plane.  

\begin{figure}[t]
	\centering
    \begin{minipage}[b]{1.0\textwidth}
    \centering
      \begin{tabular}{cc}
        \includegraphics[width=0.45\textwidth]{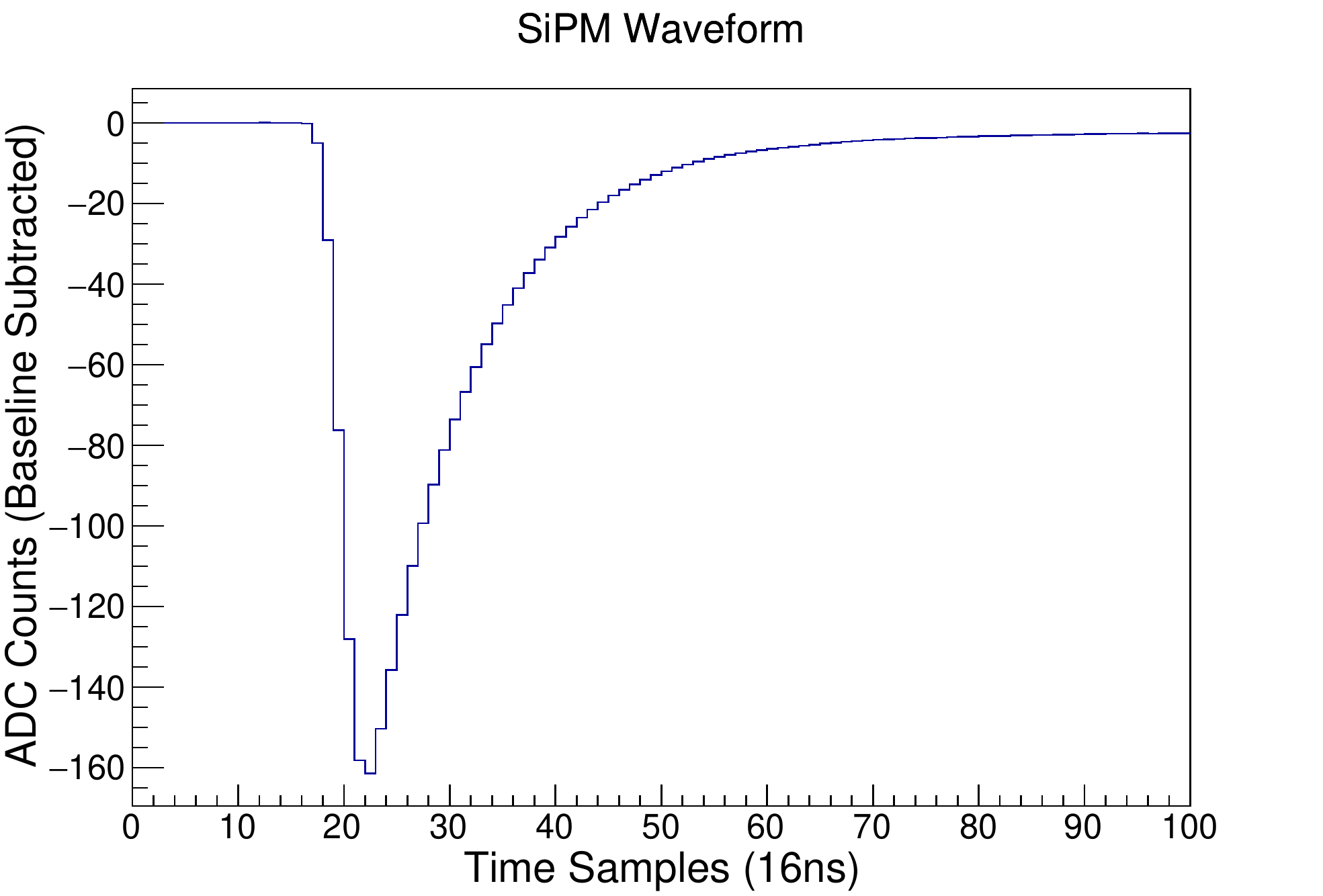} & 				\includegraphics[width=0.45\textwidth]{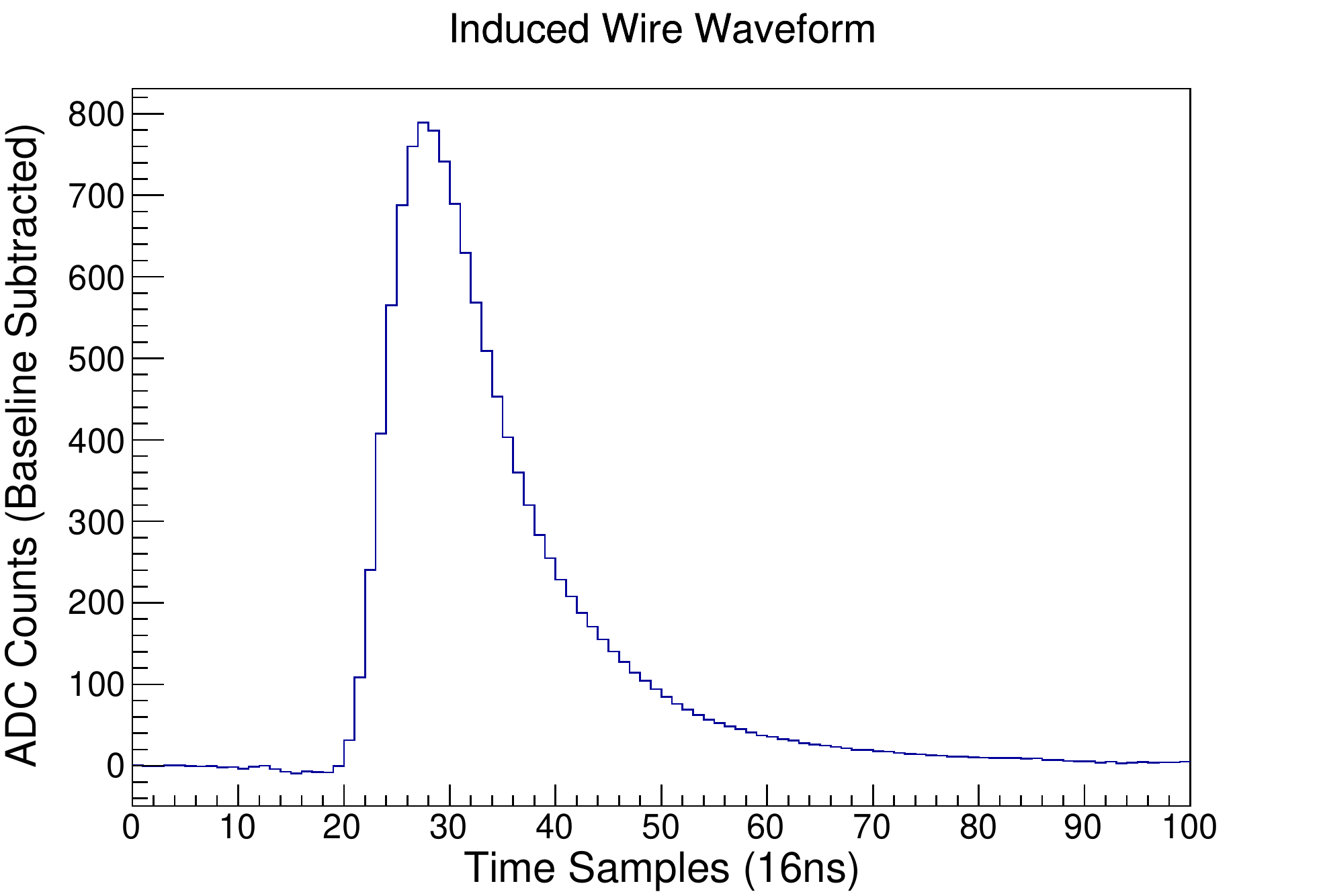} \\ 	\end{tabular}
      \caption{ Left: An inverted waveform directly from the SiPM. Right: An average amplified waveform from the wire-plane corresponding to the SiPM pulse shown to the left.}\label{fig:sipm_wire_wfm}
    \end{minipage}%
 
\end{figure}

\par

We centered a rectangular copper strip measuring 1.9 cm $\times$ 15.2 cm over one of the central wires (wire 24) and oriented it parallel to the wires. The 14 instrumented wire positions generated a distribution of induced signal amplitudes in this configuration. We also took data with the copper strip oriented perpendicular to the wires to investigate its coupling to the wire plane.  This choice of plate design was based on earlier studies of potential geometries using the MicroBooNE TPC system \cite{GabrieluBtechnote}.  The plate was placed at a vertical distance of 1.3 cm from the wire plane, and both were enclosed in a Faraday cage.  The plate was pulsed by an unamplified SensL MicroFB-SMA-60035 SiPM, that was placed in a dark box a few feet away along with an LED. The SiPM output was fed to a linear fan in/fan out module; this sent one copy of the output signal to the plate and the other to be digitized. A pulse generator fired the LED and also triggered the readout of the wire plane.

\par

\begin{figure}[tb]
	\centering
    \begin{minipage}[b]{1.0\textwidth}
    \centering
      \begin{tabular}{cc}
        \includegraphics[width=0.45\textwidth]{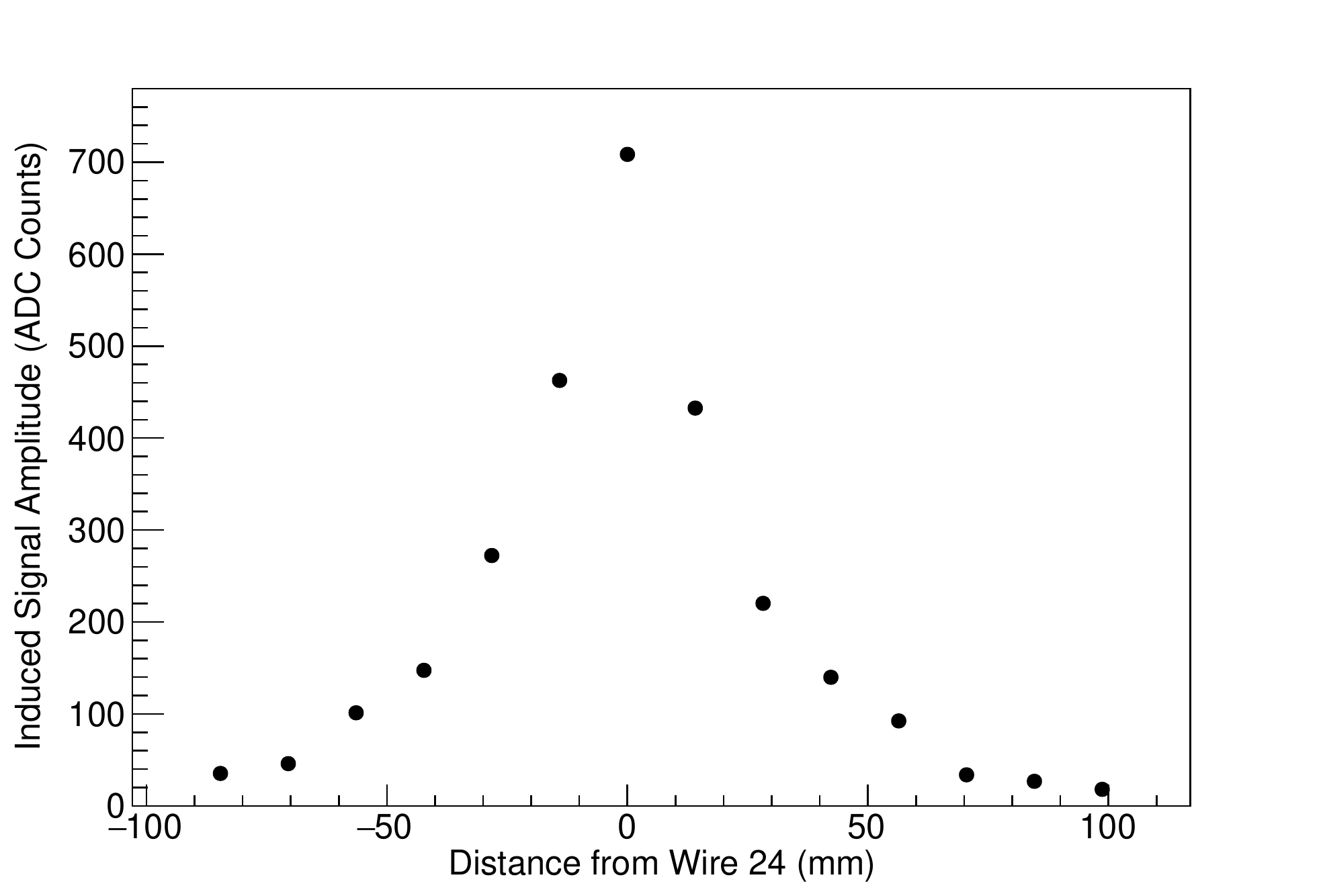}  & 				\includegraphics[width=0.45\textwidth]{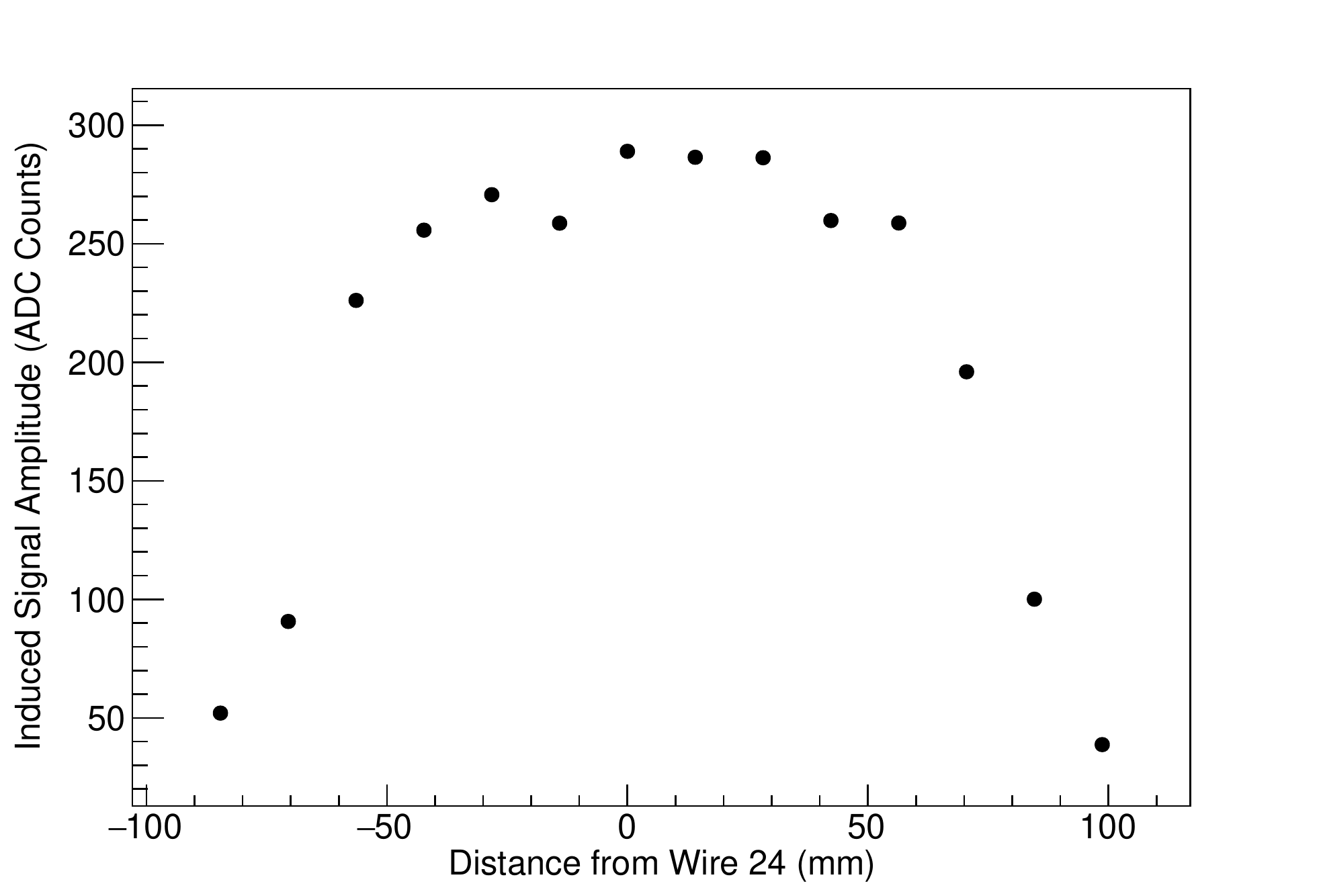}  \\ 
      \end{tabular}
      \caption{On the left, the plate is in parallel configuration 1.3 cm above wire 24. On the right, the plate is in perpendicular configuration, still 1.3 cm above the wires. The SiPM is read through its fast output in both cases. Data taken with the fast output uses a different LED amplitude than data taken with the slow output.  This should only affect only the overall normalization of the distributions shown here. 
      }\label{fig:fast_orientation}
    \end{minipage}%
 
\end{figure}

The SiPM has two output modes: the fast output, which produces a high-speed bipolar signal, and the standard output, which produces a wider, unipolar signal. All the following work uses the standard output, except for Figure~\ref{fig:fast_orientation}, which details signals from the fast output.

\par

The wires in the plane were read out through charge sensitive CAEN A1422 pre-amplifiers, designed for LArTPC systems,  with gain of 9.1 mV/fC.  These were input to a 12-bit (2 V range), 62.5 MHz CAEN DT5740 digitizer that recorded the signals, as well as output from the SiPM.  For each wire and each plate position, about 1000 waveforms were acquired and subsequently averaged to reduce the level of the ambient noise in the laboratory.  An example average SiPM pulse and the corresponding averaged wire plane signal, read out through the pre-amps and digitizer, are shown in Figure~\ref{fig:sipm_wire_wfm}.

\par

\begin{figure}[tb]
	\centering
    \begin{minipage}[b]{1.0\textwidth}
    \centering
      \begin{tabular}{cc}
        \includegraphics[width=0.45\textwidth]{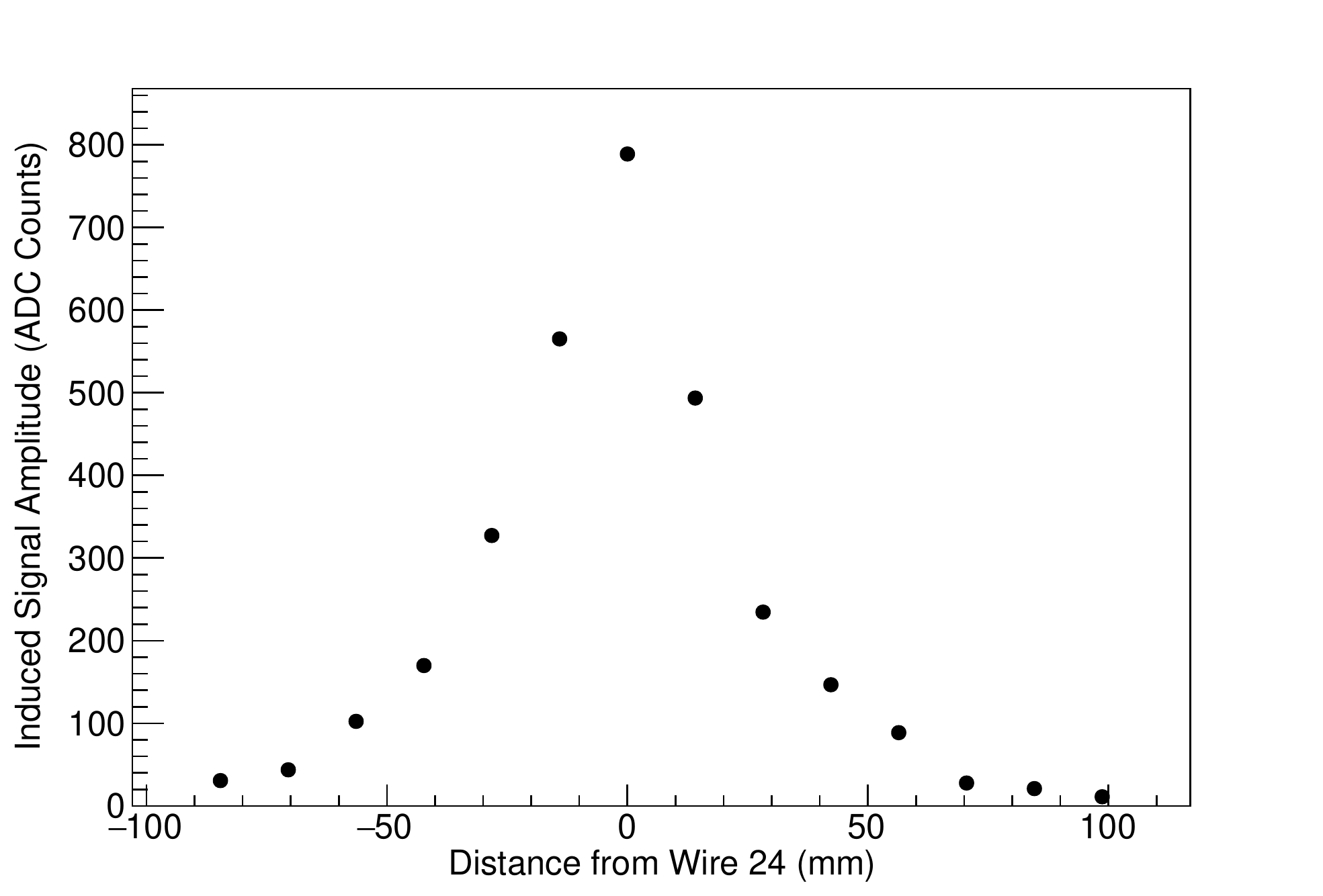}  & 				\includegraphics[width=0.45\textwidth]{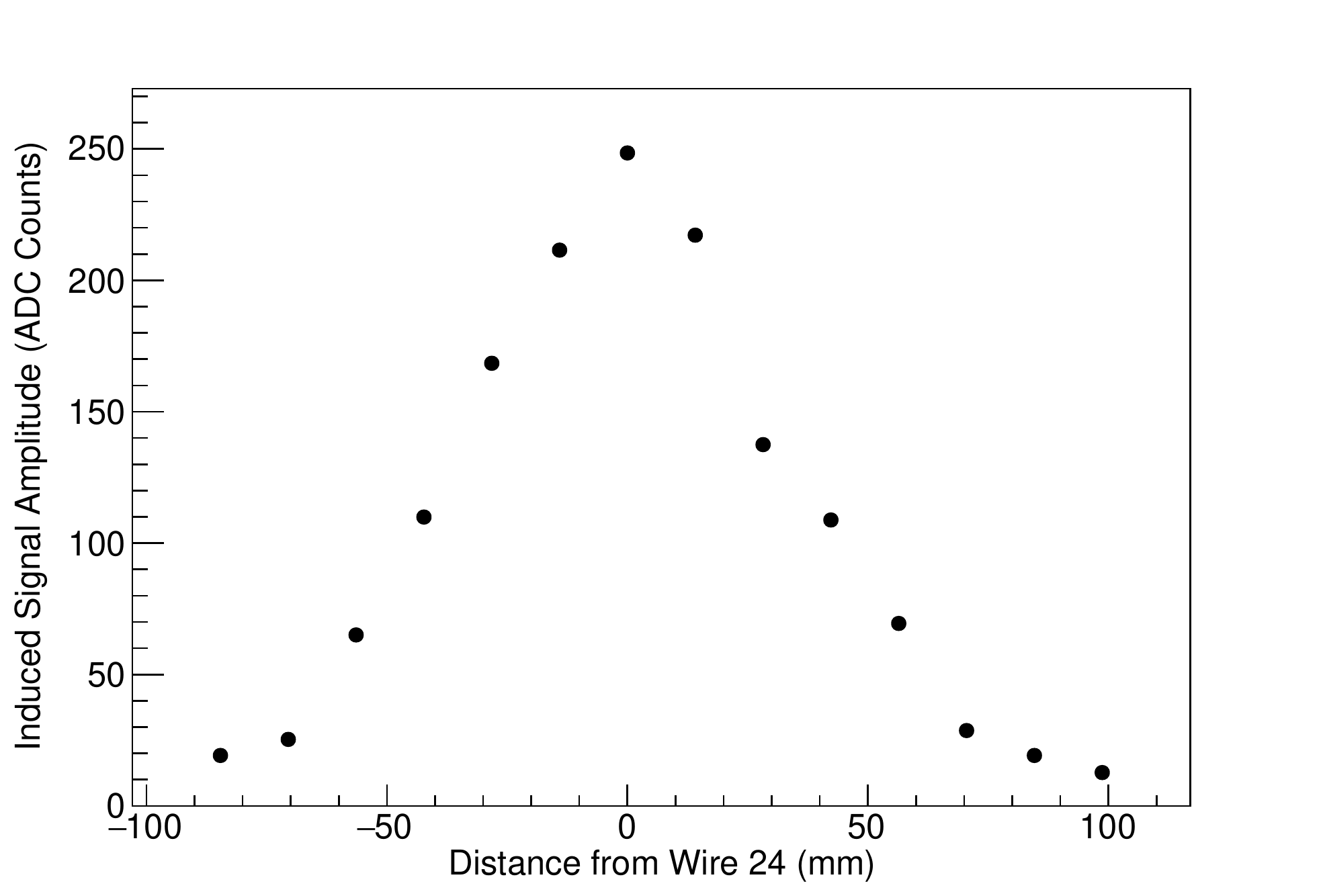}  \\ 
      \end{tabular}
      \caption{On the left, the plate is in parallel configuration 1.3 cm above wire 24. In the right, the plate is again parallel to wire 24, but it has been elevated to 3.2 cm above the wires. The SiPM is now read out through the standard output in both cases. 
      }\label{fig:slow_height}
    \end{minipage}%
 
\end{figure}

We found an amplitude distribution that is peaked around the wire over which the plate was placed (Figure~\ref{fig:fast_orientation}, left). The wire response fell off rapidly as we approached the edge of the wire plane, that may have been due to a combination of fringe fields and wire length. Due to the circular geometry of the wire-plane, wire-length decreases towards the edges. Figure~\ref{fig:fast_orientation}, right,  shows an amplitude distribution with the plate lying perpendicular to the wires at the center of the wire plane. As expected, the amplitude of the induced wire plane signal is relatively flat in the region where the area of overlap between the plate and the wires is constant, but quickly drops off at the edges due to the combined effect of shortening wires and shielding from the grounding ring pictured in Figure~\ref{fig:wireplane}. In the perpendicular configuration, the area of overlap between the plate and the wires it crosses over is greatly reduced relative to the parallel configuration. Pulse amplitudes are accordingly attenuated relative to the parallel configuration, but by a factor less than would be predicted by a naive parallel-plate capacitor model due to the presence of the fringe fields traced-out in Figure~\ref{fig:fast_orientation} (left).

Next we examined the dependence of the induced signal on the distance of the plate from the wires. Figure~\ref{fig:slow_height} (left) is a parallel plate orientation over wire 24, 1.3 cm above the wire, now using the slow output. Figure~\ref{fig:slow_height} (right) is again the product of a parallel plate configuration above wire 24, but with the plate a distance of 3.2 cm from the wire plane.  This distribution is more broadly peaked and significantly attenuated in amplitude with respect to Figure~\ref{fig:slow_height} (left).  The broad peak suggests that more fringe fields outside of the area of overlap between the plate and the wires are contributing induced signal, which explains why the attenuation of the peak is not what one would expect from a simple parallel plate capacitor model.

\begin{figure}[t] 
\centering 
\includegraphics[width=0.7\textwidth]{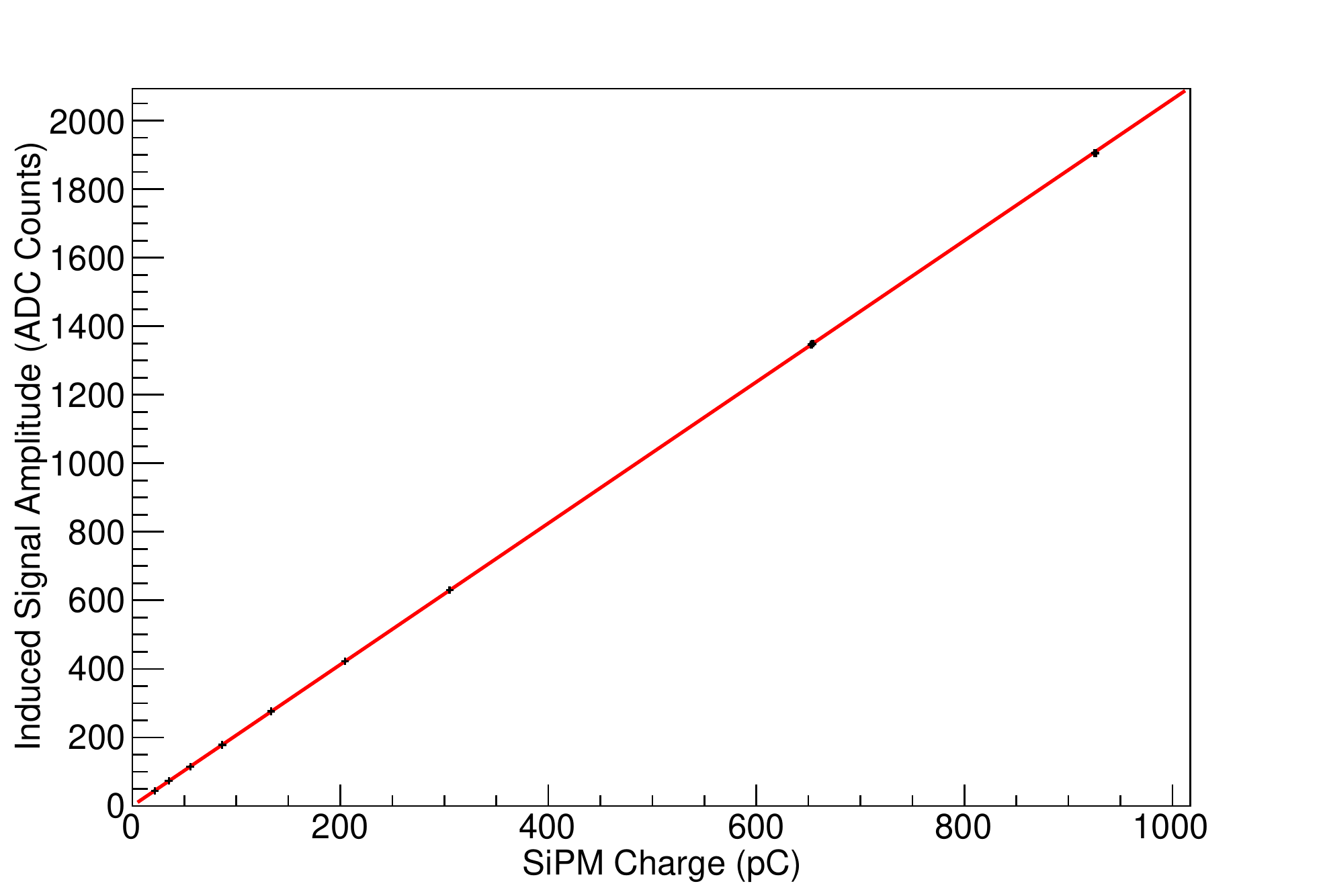} 
\caption{The signal amplitude observed on wire 24 at various SiPM charges. A linear function was fit to the data, yielding a slope of $2.063 \pm 0.002$  and a y-intercept of $-0.069\pm 0.343$ ADC counts. ${\chi}^2/\text{ndf} = 7.17/9$. Error bars are shown but occluded by point markers.}
\label{fig:linearity}
\end{figure}

By varying both the LED intensity and the SiPM bias voltage, we were able to study the dependence of the induced wire plane signal on the digitized SiPM signal.  In both cases we found a linear dependence of the induced wire plane signal amplitude on SiPM charge. An example of is shown in Figure \ref{fig:linearity}.  Based on this fit and given our assumptions above about the TPC readout, we estimate that an SPE from a 6 mm $\times$ 6 mm MicroFC-60035-SMT SensL SiPM running with a gain of $6\times10^6$ and coupled to a simple 1.9 cm $\times$ 15.2 cm plate placed 1.3 cm away from a wire plane should induce a wire plane signal of $\sim$3 ADC counts with a cold pre-amplifier gain of 14 mV/fC.

For a given cold pre-amplifier gain setting, one can boost the wire plane signal induced by an SPE by increasing the SiPM gain or plate size, or by decreasing the plate-wire plane distance.  Assuming a linear coupling of the SiPM signal to the wire plane and that the SiPM can distinguish an SPE from noise with a signal-to-noise ratio of 10:1 \cite{Denver}, one can adjust the coupling between the plate and the wire plane such that an SPE induces a signal that is roughly 10 times the noise of the TPC readout.  The linear relationship between SiPM charge and induced wire plane signal amplitudes suggests that we can extract information about the number of photoelectrons produced at each SiPM based on the induced wire plane signal.  Furthermore, as long as the plate is sufficiently close to the wire plane, the spatial distribution of induced wire plane signal amplitudes is both symmetric and tightly peaked.  These characteristics could allow us to associate wire plane signals with specific plates (and therefore SiPMs), providing information about the spatial distribution of light in the detector. This well-understood distribution will also allow us to separate signals induced by SiPMs from drift charge signals, in the event that the light signal and charge collection overlap in time in an underground detector like DUNE.

\section{Analysis}

Having demonstrated the possibility of coupling SiPM signals to the wire plane, we turn our attention to timing. Timing resolution is a subject of particular concern for wire-plane optical readout. 
Signals from wire-plane charge readouts are typically digitized at a much slower rate than signals from photodetectors. Further complicating the determination of photon arrival times is the shaping amplifier used to read out the wire-plane channels. For example, the MicroBooNE optical readout ADCs operate at a sample rate of 64 MHz on the output of shaping amplifiers with an impulse response width of 60 ns. By contrast, the MicroBooNE wire-plane readout utilizes a sample rate of 2 MHz and a shaping time of 0.5 to 3 $\mu$s. The drastic reduction in time resolution introduced by reading out PMT signals with the wire-plane readout could potentially introduce so much uncertainty into our measurement of photon arrival time that the optical system loses its utility. In what follows, we demonstrate that this is, emphatically, not the case.

\begin{figure}[tb]
	\centering
    \begin{minipage}[b]{1.0\textwidth}
    \centering
      \begin{tabular}{cc}
        \includegraphics[width=0.45\textwidth]{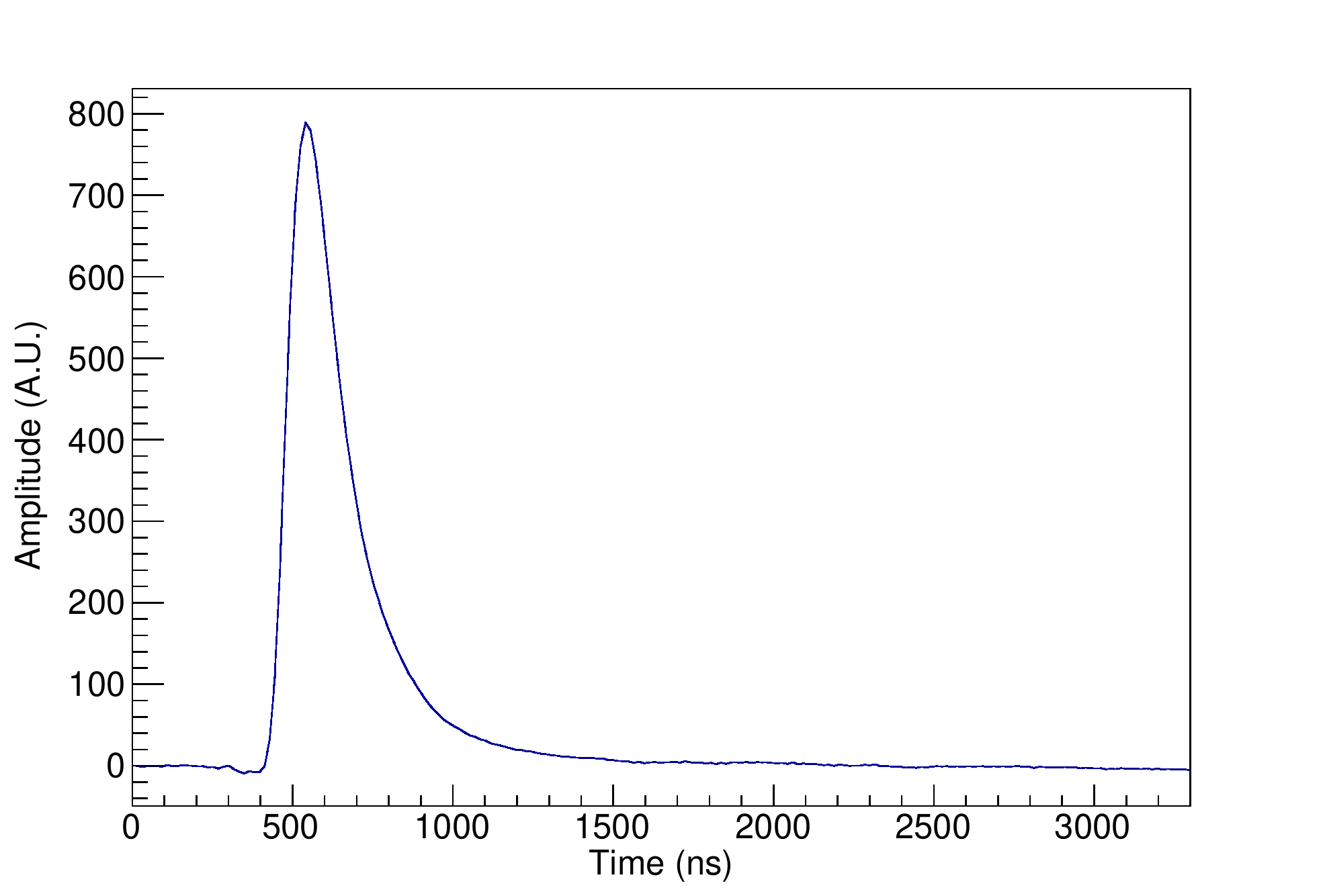}  & 				\includegraphics[width=0.45\textwidth]{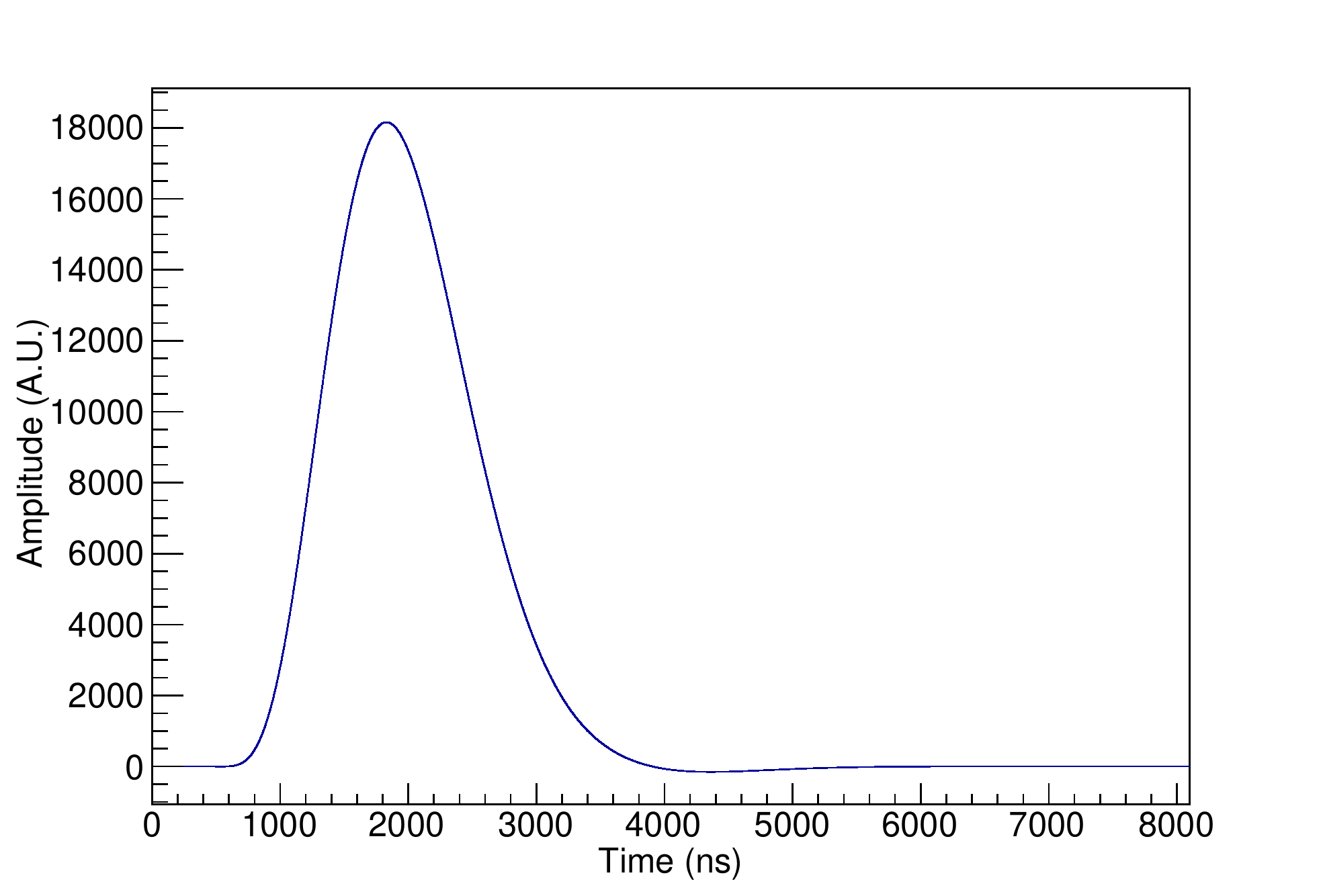}  \\ 	\end{tabular}
      \caption{ Left: The wire-plane signal, interpolated to the nanosecond. Right: The wire-plane signal after 1$\mu$s shaping.}\label{fig:signal-convolution}
    \end{minipage}%
 
\end{figure}

\begin{figure}[tb]
	\centering
    \begin{minipage}[b]{1.0\textwidth}
    \centering
      \begin{tabular}{cc}
        \includegraphics[width=0.45\textwidth]{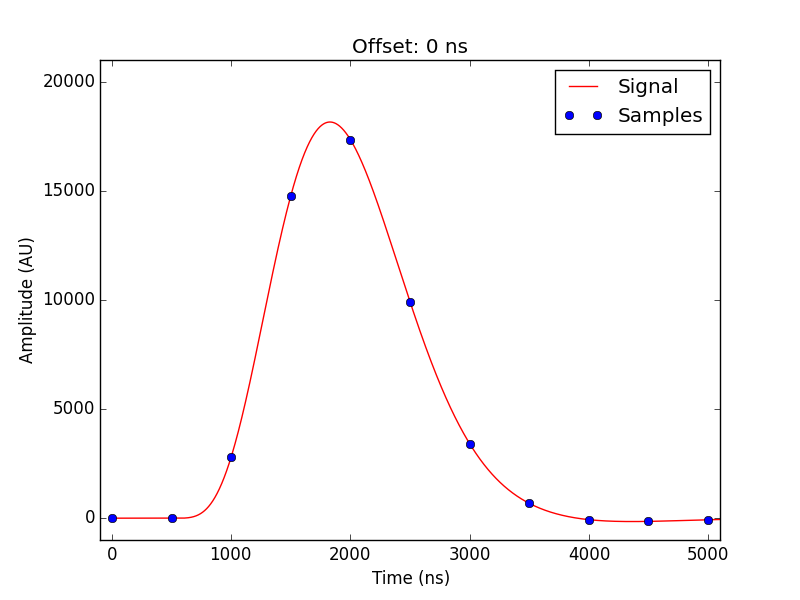}  & 				\includegraphics[width=0.45\textwidth]{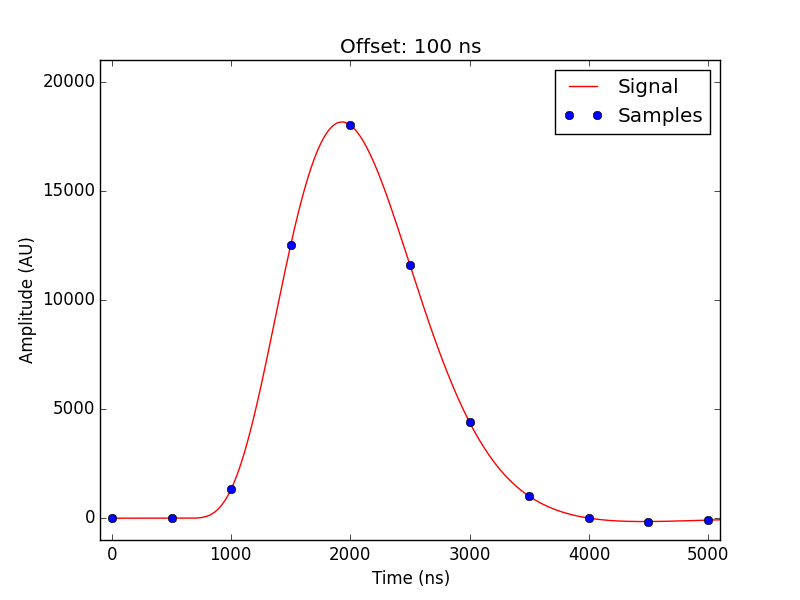}  \\ 		\includegraphics[width=0.45\textwidth]{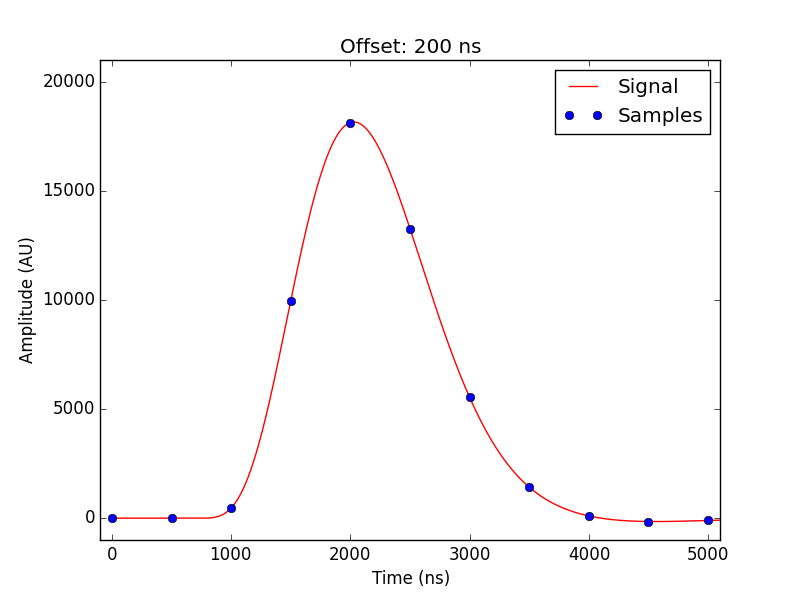}  & 			\includegraphics[width=0.45\textwidth]{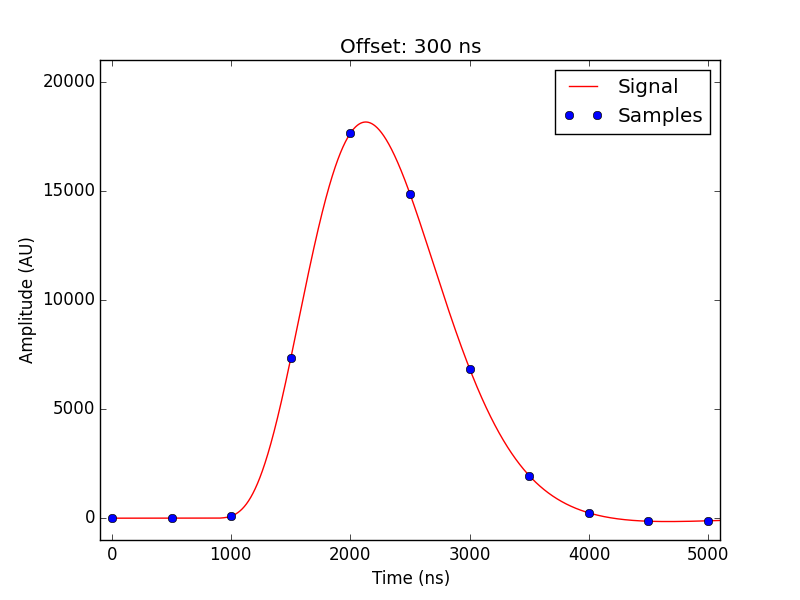} 
      \end{tabular}
      \caption{Variation in sampled waveforms driven by signal arrival time.}\label{fig:sample_variation}
    \end{minipage}%
 
\end{figure}

As a first step toward understanding our timing capabilities with a wire-plane readout system, we simulated shaping, sampling, and arrival-time reconstruction with a signal waveform from our previous measurements. These measurements were made on wire 24, which produces a large signal due to its proximity to the center of the wire-plane. The capacitive plate was placed parallel to the wires, as was the case for the measurements shown in Figure~\ref{fig:slow_height} (left). The averaged waveform was interpolated to the nanosecond, and then convolved with the impulse response of the MicroBooNE cold pre-amplifier at 1 $\mu$s shaping time. The waveform is shown before convolution in the left panel of Figure~\ref{fig:signal-convolution}, and after convolution in the right. We then proceeded to sample the shaped signal at 2 MHz at varying offsets between signal arrival time and ADC sampling clock. The sampled signal at a selection of offset times is shown in Figure~\ref{fig:sample_variation}.

\subsection{Constant Fraction Discrimination}

We extract the single photoelectron arrival time using the constant fraction discriminator (CFD) technique, which establishes a trigger based on a constant fraction of the total pulse height. The benefit of the CFD over constant threshold discriminators is the elimination of the dependence of trigger time on pulse height, assuming that the pulses in question vary only in amplitude.

\par

Our implementation of the CFD method first attenuates and inverts the signal.  It then adds the result to a delayed copy of the original signal.  The resulting bipolar waveform (the `CFD waveform') will cross zero at the constant-fraction crossing time, or ``CFD-crossing time''. This fraction can be tuned by varying delay time and attenuation fraction.  During the course of this study, we found the best performance at 80\% attenuation. All results shown here utilize a fraction of 80\% and a time delay of 500 ns, or one sample clock cycle.  

One simple approach to finding the zero crossings involves locating the maximum of the CFD waveform and following a negative gradient to a root. This method is effective, but for noisy events it can find zero crossing times that differ greatly from the zero crossing time of an analogous noiseless input pulse.  In response, we modified the zero crossing algorithm to generate a list of all zero crossings in the CFD samples, remove those with an incorrect (negative) slope, and then find the crossing closest to the maximum of the sampled signal. In the case of two equally-close samples, the earlier one is selected. After this modification, the algorithm produced more tightly clustered crossing times.

\begin{figure}[tb]
	\centering
    \begin{minipage}[b]{1.0\textwidth}
    \centering
      \begin{tabular}{cc}
       \includegraphics[width=0.45\textwidth]{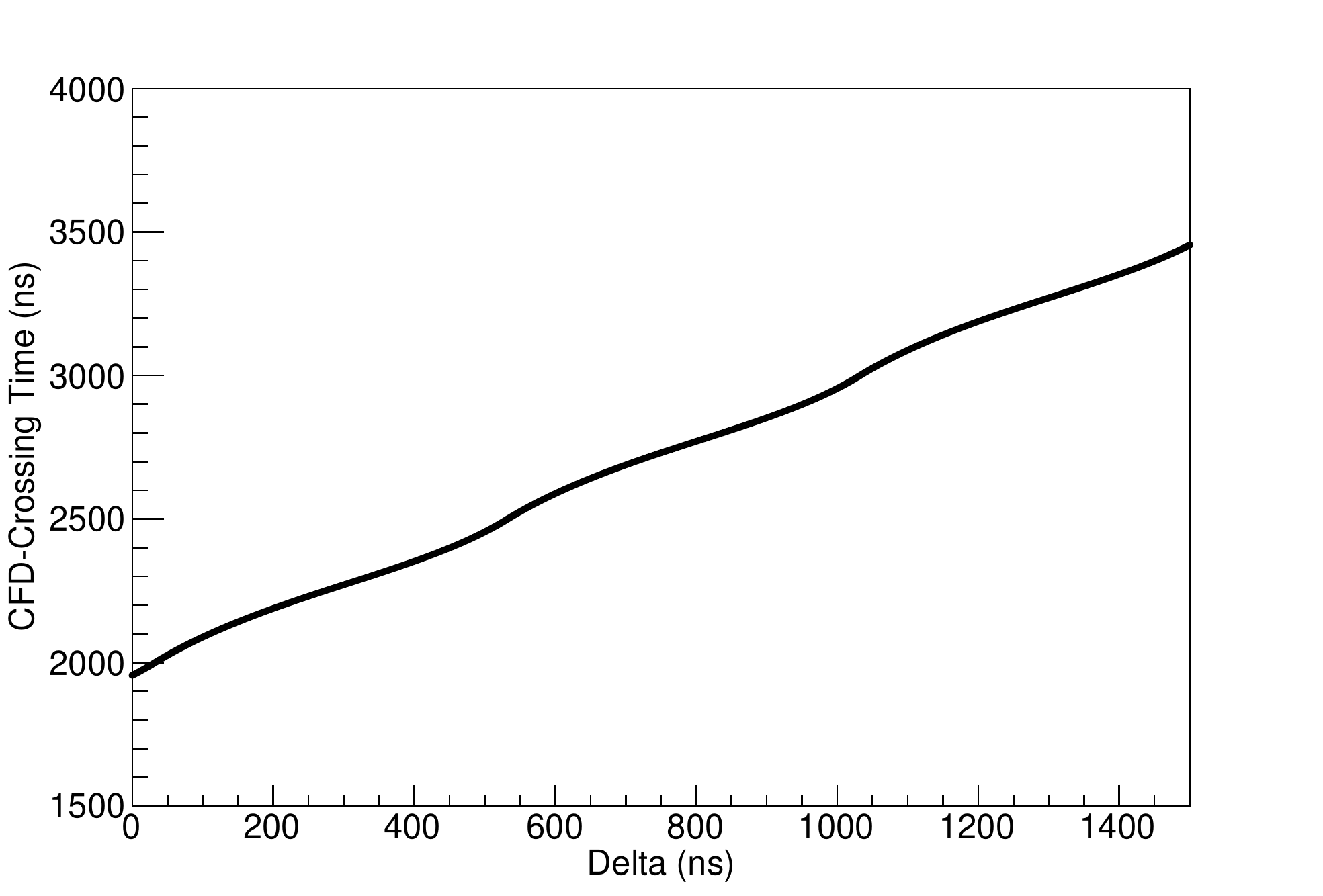} & 				                              
       \includegraphics[width=0.45\textwidth]{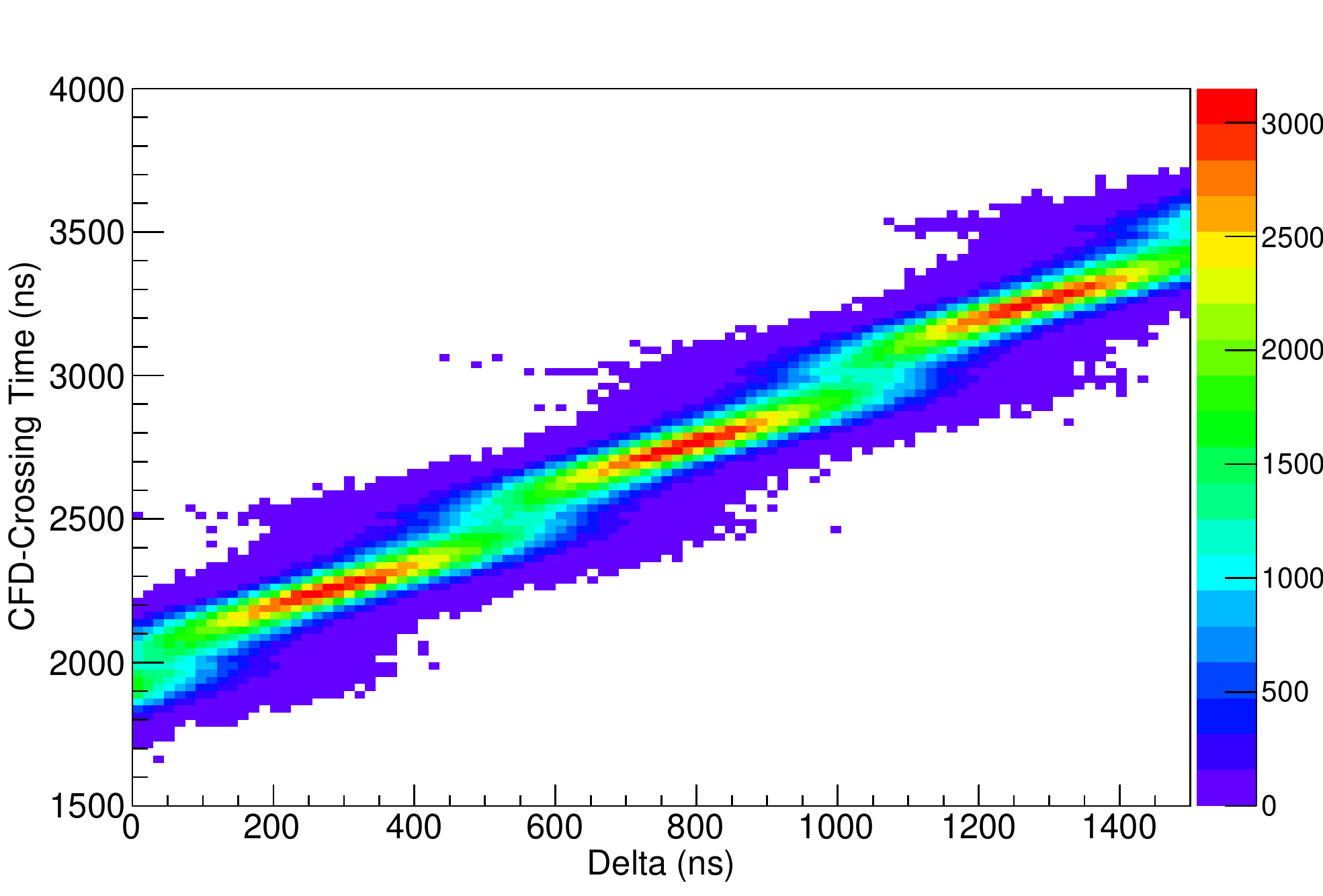}  	
        \end{tabular}
        \caption{CFD-crossing time vs. $\Delta$.  Left: without noise, for a single scan through $\Delta$.  Right: with noise, for the first 1500 ns of a 3000 ns run, representing 1000 randomly drawn noise waveforms for each value of $\Delta$.   The variations in density correspond to the changes in slope in the plot on the left.    Note the y-axis zero suppression on both plots, and that the bin width shown here is much larger than the width used for uncertainty calculations.}\label{fig:nonoise_nofilter}
    \end{minipage}%
 
\end{figure}

Given an input set of noiseless signals of equal height, the CFD method would consistently reconstruct a CFD-crossing time related both to the arrival of a photon at the photodetector and the arrival of the signal relative to the ADC sampling clock.\footnote{There is also an uncertainty corresponding to the distribution of transit times of photoelectron signals through the photodetector, which we neglect for the purposes of this study.} The delay between photon arrival time and PMT signal arrival time will be constant for all events within a given channel, so we will only consider the signal arrival time relative to our first ADC sample time. For simplicity of discussion, call this relative time $\Delta$.  To characterize the dependence of CFD-crossing time on $\Delta$, we delayed the convolved signal by 1 ns intervals from 0 to 3000 ns and sampled the signal at each iteration. Each sampled waveform was fed to the CFD algorithm, which extracted a CFD-crossing time. Figure~\ref{fig:nonoise_nofilter} (left)  shows CFD-crossing time vs. $\Delta$ from 0 to 1500 ns, where we note that the relationship between $\Delta$ and the
CFD-crossing time is one-to-one.

We then modeled the noise at the input to the wire-plane ADCs as Gaussian. After each signal delay step, we generated a noise waveform with zero mean and a standard deviation corresponding to some fraction\footnote{A 10\% noise fraction is used throughout this study, unless otherwise stated.} of the total waveform amplitude. The noise waveform was then added directly to the shaped signal before repeating the CFD procedure.

\begin{figure}[t]
\centering 
\includegraphics[width=0.7\textwidth]{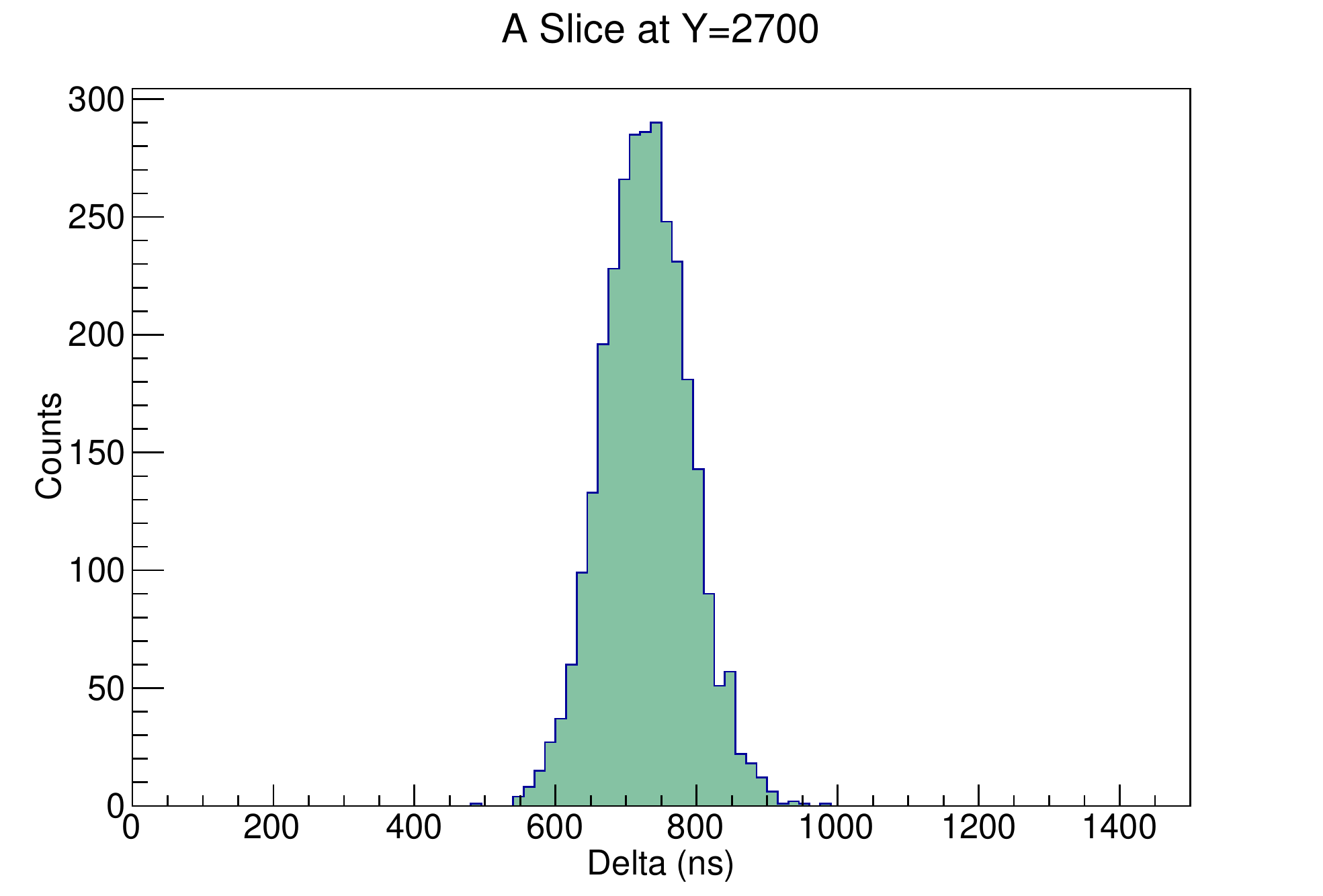} 
\caption{A slice of Figure~\protect\ref{fig:nonoise_nofilter}, right, at 2700 ns crossing time.}
\label{fig:density_slice}
\end{figure}

\begin{figure}[tb] 
\centering 
\includegraphics[width=0.7\textwidth]{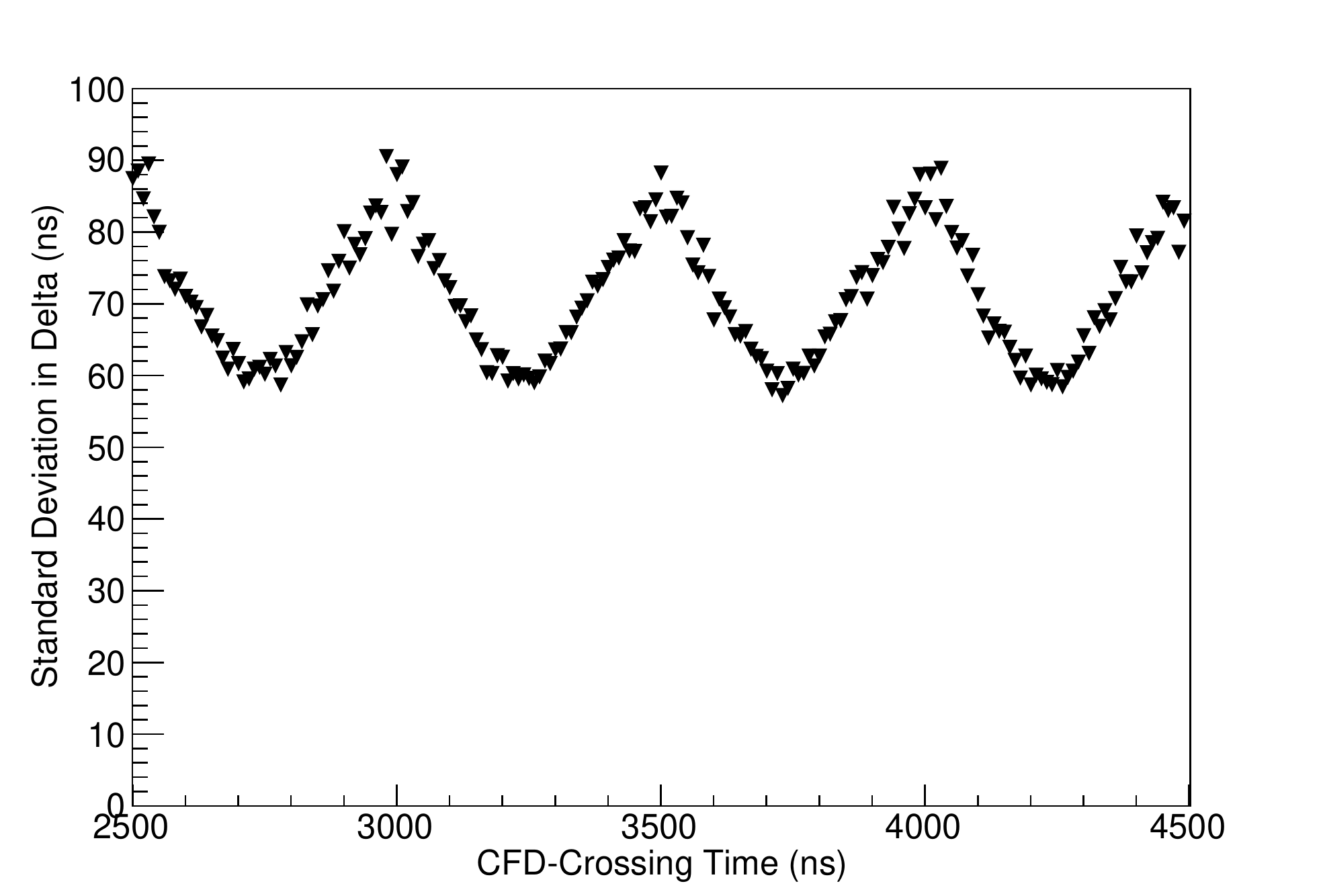} 
\caption{$\Delta$ RMS as a function of CFD-crossing time. The oscillations are due to the interplay between the CFD algorithm and the arrival time of the signal relative to the 2 MHz ADC sampling clock.}
\label{fig:rms_profile}
\end{figure}

A plot of the distribution of CFD-crossing times vs. $\Delta$ for 10\% noise is shown in Figure~\ref{fig:nonoise_nofilter} (right).  We estimated our uncertainty in reconstructing the true arrival time of the signal (in this case, $\Delta$ plus some offset) by examining distributions of $\Delta$ at each CFD-crossing time, as shown in Figure~\ref{fig:density_slice}.  The standard deviation of these distributions gives the single photoelectron timing resolution and is plotted as a function of CFD-crossing time in Figure~\ref{fig:rms_profile}.  We see a clear variation in the timing resolution due to the interplay between the CFD algorithm and the arrival time of the signal relative to the 2 MHz ADC sampling clock.

\subsection{Filtering Algorithms}

To improve on the timing resolution shown in Figure~\ref{fig:rms_profile}, we investigated the effect of applying a Wiener filter to the signal before applying the CFD algorithm.   In frequency space, the Wiener filter takes the form: 

\begin{equation}
W(f) = \frac{S^2(f)}{S^2(f)+N^2(f)}
\label{eqn:wiener_freq}
\end{equation}

\begin{figure}[tb] 
\centering 
\includegraphics[width=0.7\textwidth]{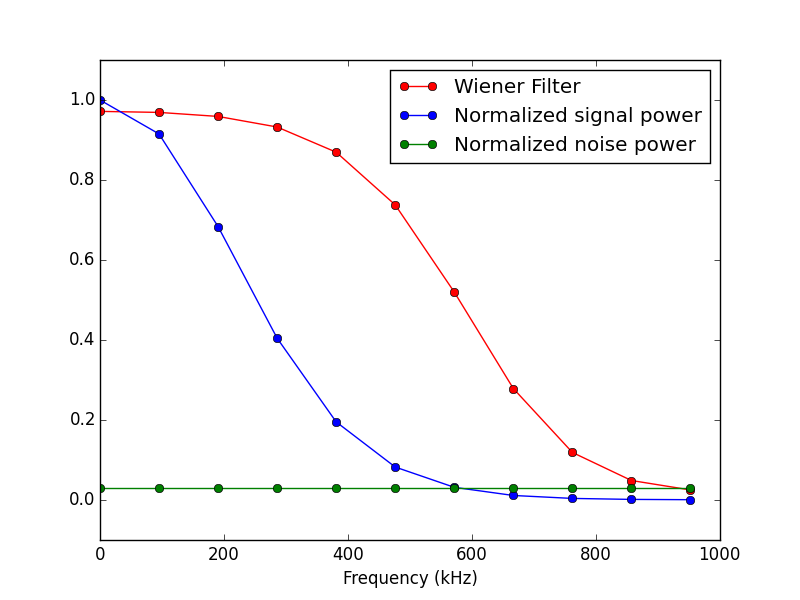} 
\caption{The average power spectrum of the signal averaged over the 500 $\Delta$, as well as a noise power spectrum at 10\% noise fraction averaged over 500 runs. Both spectra are normalized to the maximum of the signal spectrum. The Weiner filter constructed from these power spectra according to equation~\protect\ref{eqn:wiener_freq} is also shown.}%
\label{fig:wiener_filter}
\end{figure}

The form of the filter relies only on the power spectra of the signal and the noise. We have approximated our noise as white, so $N^2(f)$ is constant and easily calculated. We estimate $S^2(f)$ by using the average power spectrum for 500 different values of $\Delta$ chosen in 1 ns increments from 1 to 500 ns.  The power spectra and corresponding Wiener filter are shown in Figure~\ref{fig:wiener_filter}.

\begin{figure}[tb]
	\centering
    \begin{minipage}[b]{1.0\textwidth}
    \centering
      \begin{tabular}{cc}
\includegraphics[width=0.45\textwidth]{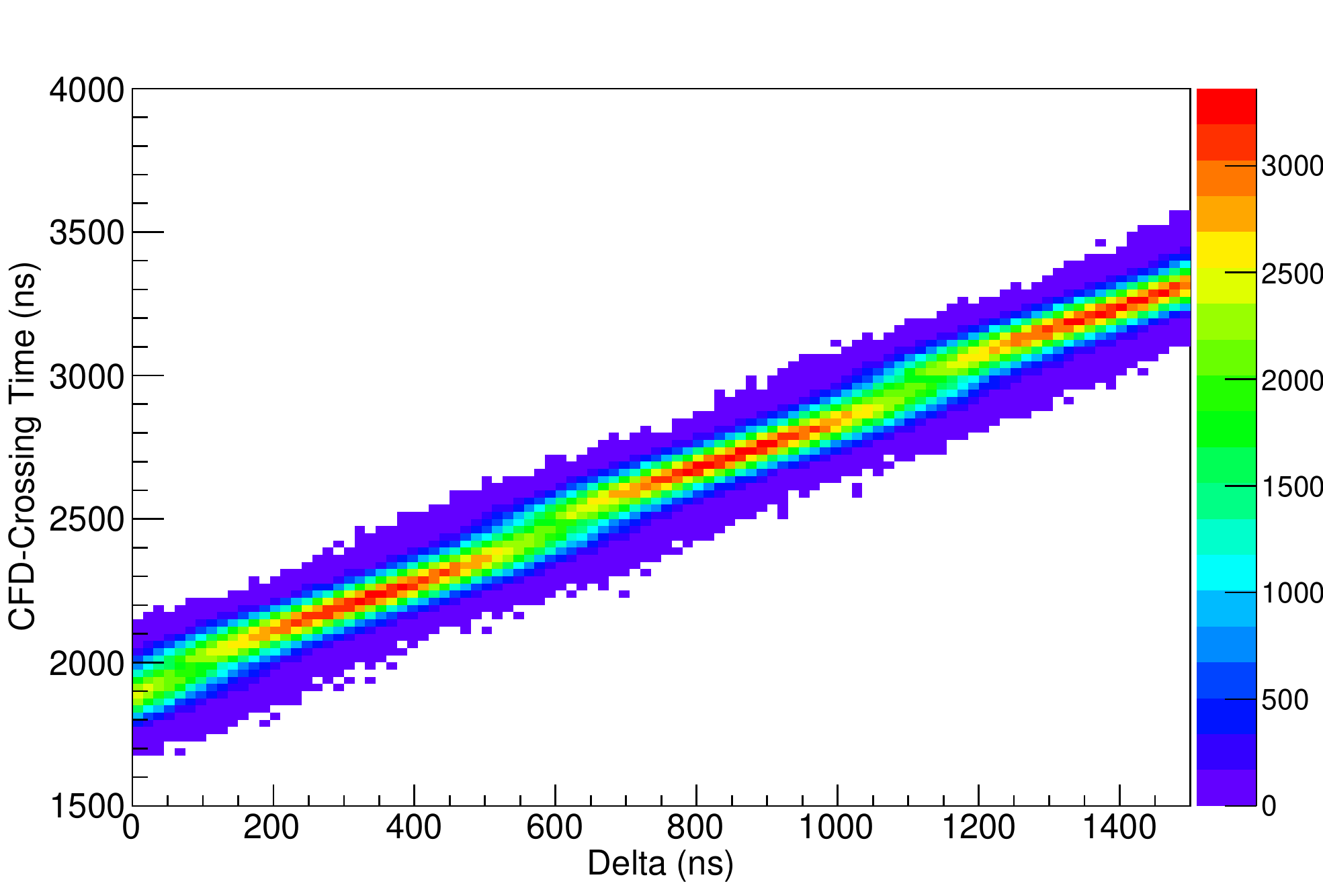} & 				\includegraphics[width=0.45\textwidth]{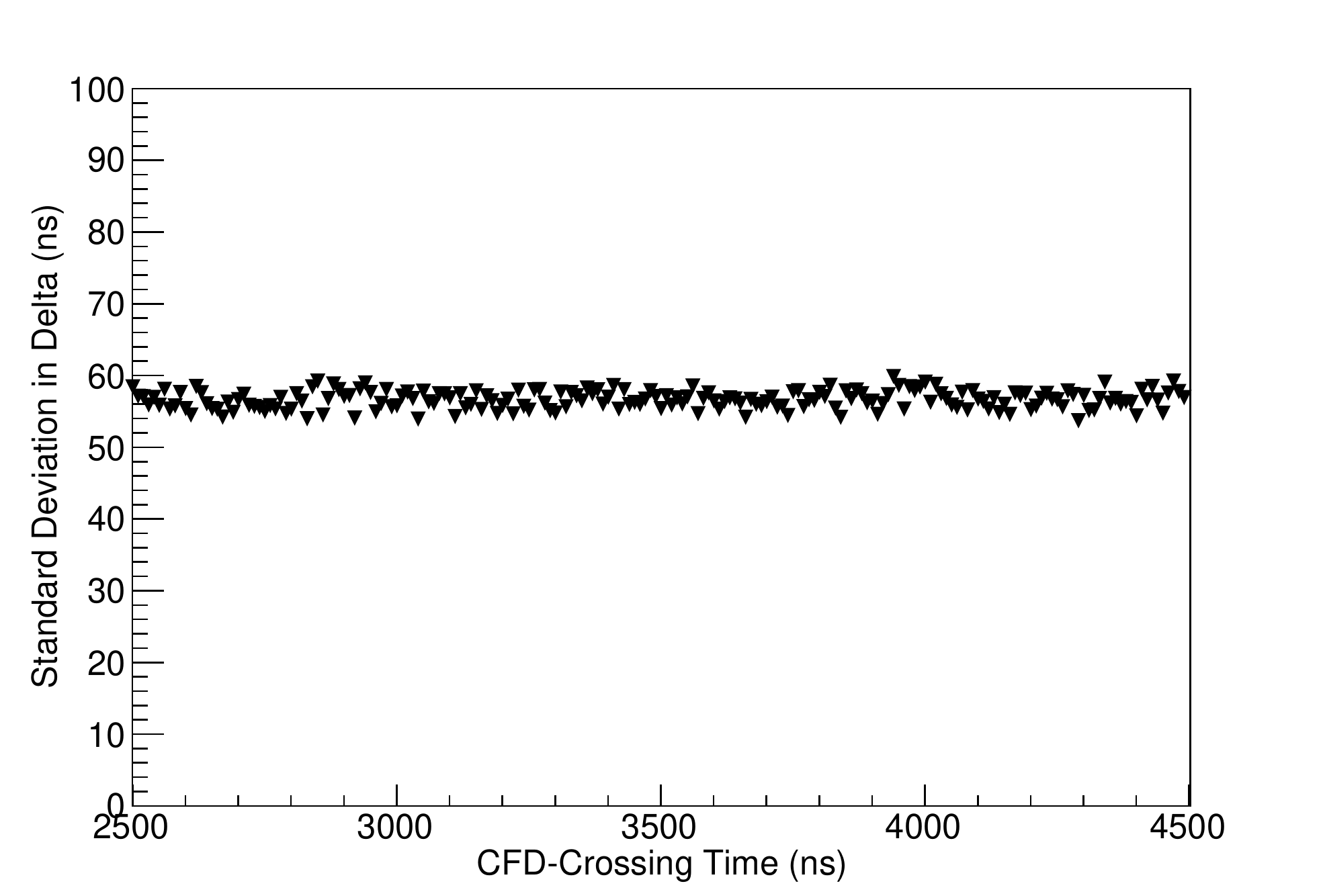} \\ 	\end{tabular}
      \caption{Left: CFD-crossing time vs. $\Delta$ after the application of the Wiener filter to reduce noise. As before, these are the first 1500 ns of a 3000 ns run, representing 1000 randomly drawn noise waveforms for each value of $\Delta$. Left: standard deviation in $\Delta$ as a function of CFD-crossing time after application of the Wiener filter.}\label{fig:rmsfilters}
    \end{minipage}%
 
\end{figure}

We then applied this Wiener filter to each of our waveforms generated with 10\% noise and re-computed the CFD-crossing times at each value of $\Delta$.  The result is shown in Figure~\ref{fig:rmsfilters} (left).  Again, by computing the standard deviation of the distribution of $\Delta$ values for each CFD-crossing time, we can compute the corresponding single photoelectron timing resolutions, shown in Figure~\ref{fig:rmsfilters} (right).   Not only does the Weiner filter reduce the timing resolution to just under 60 ns, but it also removes the dependence of the timing resolution on the arrival time of the signal relative to the 2 MHz ADC sampling clock.

\subsection{Exploiting Adjacent Wires}

So far, we have only considered the induced signal on the wire immediately beneath the plate. This is the wire that will carry the strongest signal, but as evidenced by Figure~\ref{fig:slow_height} (left), there are significant signals induced on nearby wires due to fringe fields. Provided the signal amplitude doesn't drop off too quickly, we may be able to improve our signal-to-noise ratio by summing the signals of the maximal wire and some of its adjacent wires. If we assume that the noise is uncorrelated wire-to-wire, then summing $n$ wires will give a waveform whose noise is characterized by a standard deviation $\sigma_n$ that is equal to $\sqrt{n}\cdot\sigma_1$, where $\sigma_1$ is the standard deviation of the noise on a single wire.

Let $A_1$ be the amplitude of the signal immediately under the plate. Let $f(x)$ denote the fractional amplitude of the fringe field at some distance $x$ from the wire directly under the plate. $f(x)$ is normalized to a maximum value of 1 at $x=0$, immediately under the capacitive plate. The total signal amplitude ($A_n$) will be:
\begin{equation}
A_n = A_1 \cdot \sum_{j=-n/2}^{n/2} f(\delta_x \cdot j)
\label{eqn:signal_strength}
\end{equation}
where $\delta_x$ is the wire spacing, or ``pitch''.

\begin{figure}[tb] 
\centering 
\includegraphics[width=0.7\textwidth]{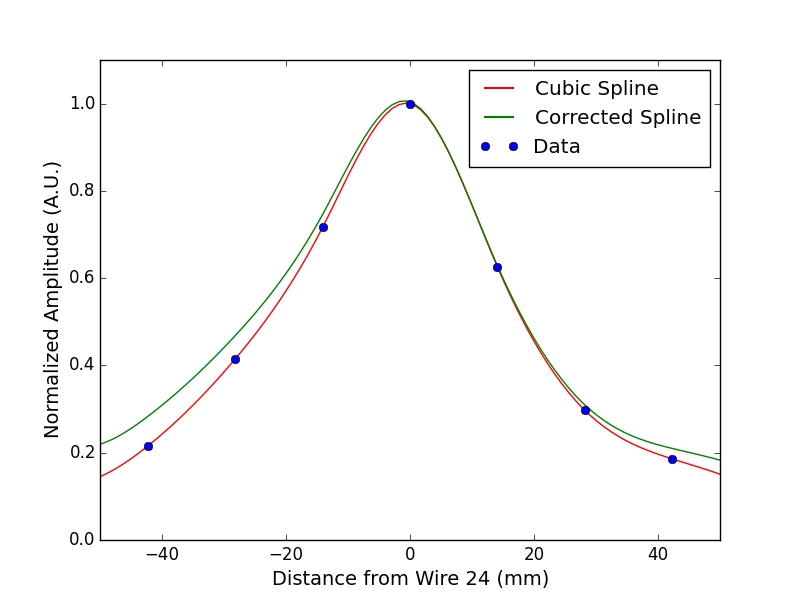} 
\caption{A cubic spline fit to the data from our wire-plane experiments. The curve in green is the fit spline with a correction for wire length dependence.}
\label{fig:falloff}
\end{figure}

\begin{figure}[t] 
\centering 
\includegraphics[width=0.7\textwidth]{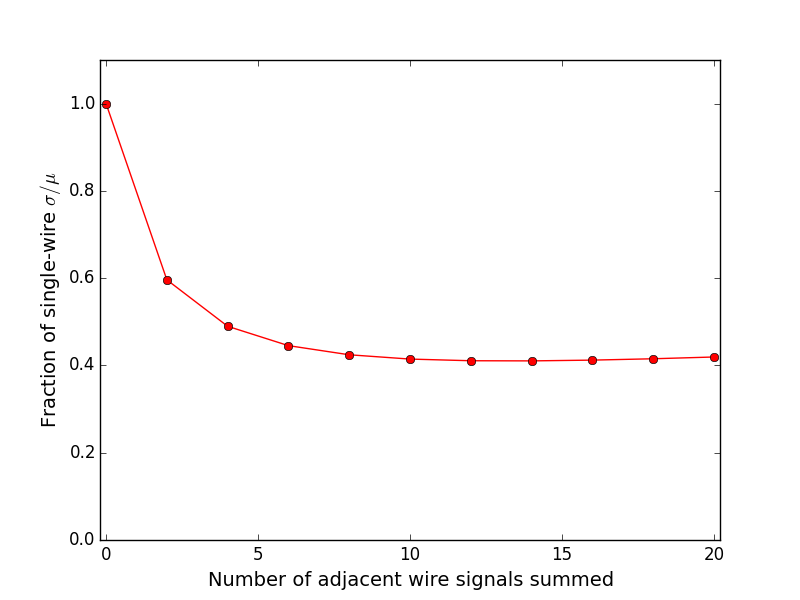} 
\caption{The reduction in noise fraction as a function of the number of adjacent wires summed with the wire immediately beneath the capacitive plate.  The function quickly approaches a minimum, and then begins to grow as the signal dies off.}
\label{fig:wire_count}
\end{figure}

The experiments described in Section~\ref{Experiments} used a circular wire-plane with a 4.7 mm pitch. However, measurements were only made with every third wire, giving an effective spacing of 14.1 mm.  In order to determine the effect of adding the signals from adjacent wires with a few mm spacing we interpolated the measured distribution in Figure~\ref{fig:slow_height} between $\pm50~$mm. We chose a simple cubic spline interpolation, and then estimated the fractional amplitude of the fringe field $f(x)$ by correcting for the difference in wire length directly beneath our plate and at a distance $x$ away.  The cubic spline as well as the corrected spline taking into account wire length is shown in Figure~\ref{fig:falloff}.  Assuming a wire pitch of 4.7 mm, we can compute then $A_n$ as a function of the number of adjacent wires included in the summation.  The corresponding reduction in the noise relative to the signal is plotted in Figure~\ref{fig:wire_count} as a function of $n$.

\begin{figure}[t] 
\centering 
\includegraphics[width=0.7\textwidth]{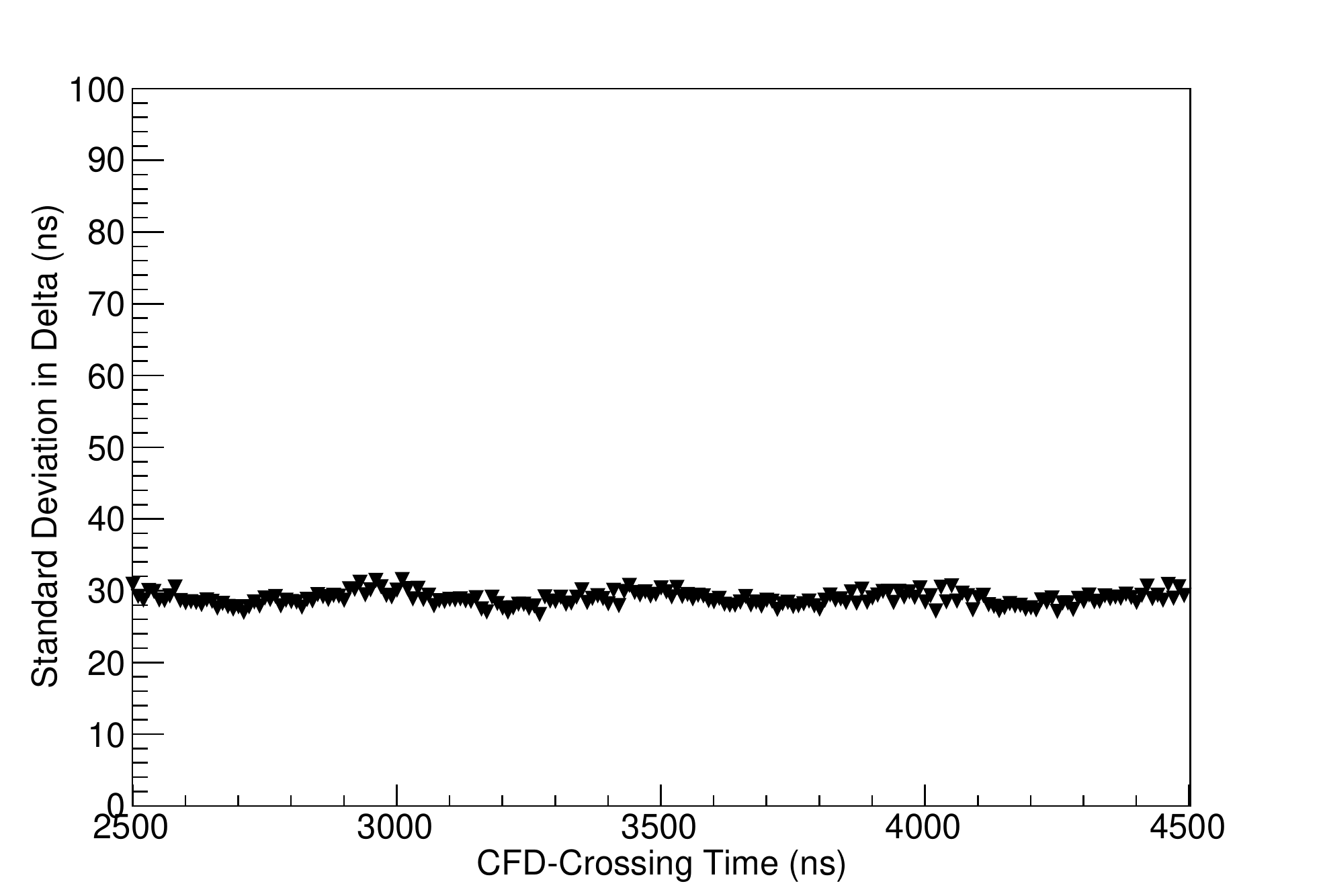} 
\caption{Standard deviation in $\Delta$ as a function of CFD-crossing time. With the addition of signal from four adjacent wires, the uncertainty in $\Delta$ from the CFD is now $\sim$30 ns for all values of $\Delta$.}
\label{fig:wiener_halfnoise_rms}
\end{figure}

Using only 4 adjacent wires, 
we can achieve an increase in the signal-to-noise ratio of just over a factor of 2. To quantify this in terms of timing resolution, the Wiener-filtered simulation was repeated with a 5\% noise fraction (one half of the standard 10\%), yielding the $\Delta$ standard deviation values plotted in Figure~\ref{fig:wiener_halfnoise_rms}. The distribution now has a mean value of $\sim$30 ns, and, as before, does not depend on the arrival time of the signal relative to the 2 MHz ADC sampling clock.

\section{Potential Implementation}
\par

In ultra-large, underground LArTPC detectors such as the single-phase DUNE design, a thin profile light collection system that can fit between adjacent wire planes minimizes the dead volume between drift regions.  A detector consisting of wavelength-shifting light guides whose ends are read out by small photodetectors (e.g. SiPMs) has been proposed as a cost-effective realization of such a system~\cite{Demonstration}.   The SiPMs attached to these light guides could be read out by the design we propose here.  

\par
Alternatively, one could also imagine designing a system consisting only of SiPMs coated with wavelength shifter and no light guides (also under development are 128 nm sensitive SiPMs that would not require coating with wavelength shifter \cite{Hamamatsu}).  There are two competing effects when comparing the light guides to a SiPM-only system, that can be tuned to reach equivalent light collection efficiency.   First, considering only the active region,  the average number of photoelectrons detected per incident VUV photon will be much lower for a light guide bar system than for a SiPM-only system.  On the other hand, the active area of light guide bar is large compared to that of a single SiPM.  Unfortunately, such a SiPM-only system would be impractical with a dedicated readout, because this design requires many times more SiPMs than are used in the light guide bar system to obtain the same efficiency-weighted active area.   However, the anode-coupled readout design described above makes this system potentially practical.

\par
As a specific example, consider 4 $\times$ 4 arrays of 6 mm $\times$ 6 mm SiPMs.  These could be connected to plates similar in size to the one studied in Sec.~\ref{Experiments}, positioned adjacent to the TPC wires.  The coupling between the plates and the TPC wires should be adjusted such that the baseline noise induced by the SiPM on the wires is less than the noise already present on the wires.  One can then stagger the lateral positions of the plates such that one can unambiguously correlate an induced signal to one of many plates (and therefore SiPM arrays) coupled to the same wire plane. 

\par
There are a few complications associated with coupling a SiPM array to the TPC wires that are foreseen.  For example, the increased capacitance of the array relative to a single SiPM may slightly degrade the single photoelectron timing resolution, though this is likely a small effect~\cite{array1,array2} relative to the 1 $\mu$s shaping and 2 MHz sampling of the signal.  Furthermore, the dark rate of these arrays is on the order of $\sim$10 Hz/SiPM \cite{DUNEdesign}, which implies that each plate has a $\sim$30\% probability of inducing a signal every 2.3 ms.  However, the duration of these induced signals would be driven by the shaping time of the TPC readout ($\sim1 \mu$s), which is small in comparison to the full 2.3 ms readout window in DUNE.

\par
Regardless of the light collection design, the anode-coupled readout is compatible with any triggering plan that incorporates TPC information since the induced signals produce a characteristic pattern across very specific wires.  Pattern recognition algorithms can be implemented in FPGAs or higher level software to form triggers from the expected wire signals.  This is, for example, consistent with the present plan for DUNE triggering \cite{DUNEdesign}.

\par Some TPC wire chamber designs, including DUNE, introduce a uninstrumented grounded ``mesh,'' located directly behind the collection plane \cite{DUNEdesign}.    In fact,  the original MicroBooNE design called for such a mesh \cite{uBTDR},  but this was eliminated during installation to reduce complexity.   Assuming charge collection in MicroBooNE is demonstrated to be successful without the mesh, then it is likely that the mesh will be dropped in future detector wire chambers, reducing complexity and cost.   In this case, our proposed design is directly applicable.  If a mesh is included, then the capacitive plates can be mounted on the grounded mesh, facing the collection wires.

\par
There is ample space for installation of capacitive plates adjacent to photodetectors in a typical LArTPC detector, since these light collection systems generally have sparse geometric coverage.  The geometric coverage presently planned for the DUNE experiment is 0.2\% \cite{DUNEdesign}.  The geometric coverage for MicroBooNE is 0.9\%, while the coverage for the ICARUS and CAPTAIN detectors is 0.5\% \cite{MichelSorel}.  Even if the coverage is greatly increased because of the cost savings of the anode-coupled readout, space for implementing the capacitive plates will not be an issue.

\section{Discussion}

We have demonstrated the sensitivity of a wire-plane to unamplified SiPM signals coupled via capacitive plates. Extrapolations based on the measured linearity of the response show that this design is sensitive to single photoelectron pulses. If the coupling plates are situated close to the wire-plane and far from its edges, the amplitude distribution of induced signals over the wire-plane is large, symmetric, and tightly peaked. It may be possible to use this characteristic shape to efficiently distinguish SiPM signals from true charge signals. The long shaping times and low frequency ADC clocks associated with wire-plane readouts present a challenge to accurate photon arrival time determination. Nevertheless, using realistic parameters from MicroBooNE's wire-plane readout, we have shown that the combined application of a constant-fraction discriminator and a Wiener filter reduces our single photoelectron timing uncertainty to $\sim$60 ns.  If we additionally exploit the signals induced on adjacent wires, we can further reduce the uncertainty to $\sim$30 ns.    
\par

This work opens up future interesting studies for ultra-large LArTPCs.  Although not required in these tests, amplification of the signal into the plate might be motivated for some LArTPC applications.   Further study of the separation of the light-induced signal from the charge induced signal would show whether this design can be employed in near-surface and on-surface LArTPC detectors.   Construction of a prototype system that can run in an existing, well-shielded, small LArTPC could allow measurement of rates of and study of the signal from $^{39}$Ar decays that may need to be subtracted in the case of ultra-large LArTPC detectors.  Such a study would also provide the next level of demonstration of the design, as is needed for prototyping.   Also, this prototype could be used to demonstrate that the two time constants of the argon scintillation light, 6 ns and 1.6 $\mu$s,  can be separated, demonstrating the particle identification capability \cite{pid} of this readout system.

\par

This readout system may be applicable to low-rate TPCs where high radiopurity is required.  Examples may be neutrinoless double beta decay detectors using a design similar to EXO \cite{EXO} or a future neutrino magnetic moment search using a design similar to MuNu \cite{munu}.    In these detectors,
cables that can carry radioactive contaminants can be replaced with a capacitive plate made out of radiopure copper \cite{radiopure}. 

\par

In summary, the work presented in this paper motivates consideration of employing this design in future LArTPCs such as DUNE.  The multiple benefits are reduced cabling and feed-throughs, hence reduced impurities and heat-leaks,  and reduced costs.  Removing these limitations could allow for an expanded and improved light collection detector.  Excellent performance has been demonstrated by this study, indicating that this is a viable and valuable alternative design to the present dedicated light collection readout that should be considered for the future.

\section*{Acknowledgements}

The authors thank the MicroBooNE collaboration for allowing initial studies of use of plates with a TPC.  While those results are not reported here, these data were valuable to the development of the concept and the choice of the plate design.   We thank S. Pordes of Fermilab for the loan of the LArTPC wire plane for the tests described in Section~\ref{Experiments}.  We thank J. Spitz, L. Winslow and T. Wongjirad for useful comments throughout the studies.  We thank J. Asaadi and T. Strauss for reading of an early draft.    The authors are funded by NSF grant PHY-1505855.

\end{document}